\documentclass[sigconf,screen]{acmart}

\usepackage[T1]{fontenc}
\usepackage[utf8]{inputenc}
\usepackage{microtype}
\usepackage{xspace}
\usepackage{amsmath}
\usepackage{graphicx}
\usepackage{textcomp}
\usepackage{xcolor}
\usepackage{soul}
\usepackage{url}
\usepackage{paralist}
\usepackage[inline]{enumitem}
\usepackage{colortbl} 
\usepackage{cleveref}
\usepackage{listings} 
\usepackage{pifont}
\usepackage{multirow}
\usepackage{booktabs}
\usepackage{adjustbox}

\usepackage{tcolorbox}
\usepackage{tabularx}

\def\BibTeX{{\rm B\kern-.05em{\sc i\kern-.025em b}\kern-.08em
    T\kern-.1667em\lower.7ex\hbox{E}\kern-.125emX}}

\newcommand{\para}[1]{%
  \noindent\textbf{#1.\ }%
}

\crefformat{section}{§#2#1#3}
\Crefformat{section}{§#2#1#3}

\definecolor{dollarbill}{rgb}{0.52, 0.73, 0.4}
\definecolor{beaublue}{rgb}{0.74, 0.83, 0.9}
\definecolor{ao(english)}{rgb}{0.0, 0.5, 0.0}
\definecolor{amaranth}{rgb}{0.9, 0.17, 0.31}
\definecolor{cerublue}{rgb}{0.16, 0.32, 0.75}
\definecolor{frenchblue}{rgb}{0.0, 0.45, 0.73}
\definecolor{iceberg}{rgb}{0.44, 0.65, 0.82}
\definecolor{blue-violet}{rgb}{0.24, 0.17, 0.99}
\definecolor{darkorchid}{rgb}{0.6, 0.2, 0.8}
\definecolor{my-violet}{rgb}{0.80,0.79,1.00}
\definecolor{ao}{rgb}{0.0, 0.0, 1.0}
\definecolor{atomictangerine}{rgb}{1.0, 0.6, 0.4}
\definecolor{alizarin}{rgb}{0.82, 0.1, 0.26}
\definecolor{americanrose}{rgb}{1.0, 0.01, 0.24}
\definecolor{amber}{rgb}{1.0, 0.75, 0.0}
\definecolor{amber(sae/ece)}{rgb}{1.0, 0.49, 0.0}

\newtcolorbox{quotebox}{colback=dollarbill!30,boxrule=0.4pt,colframe=black,fonttitle=\bfseries,top=2pt,bottom=2pt, before skip=2pt, after skip=2pt, left=2pt,
right=2pt,
top=2pt,
bottom=2pt}

\newcommand{\toolName}{\texttt{SkillMOO}}

\setcopyright{cc}
\setcctype{by}
\acmYear{2026}
\copyrightyear{2026}
\acmDOI{10.1145/3832783.3834550}
\acmISBN{979-8-4007-2882-2/2026/10}
\acmConference[ASE '26]{Proceedings of the 41st IEEE/ACM International Conference on Automated Software Engineering}{October 12--16, 2026}{Munich, Germany}
\acmBooktitle{Proceedings of the 41st IEEE/ACM International Conference on Automated Software Engineering (ASE '26), October 12--16, 2026, Munich, Germany}
\acmSubmissionID{ase26nier-p65-p}
\received{2026-05-13}
\received[accepted]{2026-07-02}

\begin{document}

\title{SkillMOO: Multi-objective Optimization of Agent Skills for Software Engineering}

\author{Jingzhi Gong}
\correspondingauthor
\orcid{0000-0003-4551-0701}
\affiliation{%
  \institution{King's College London}
  \city{London}
  \country{United Kingdom}
}
\email{jingzhi.gong@kcl.ac.uk}

\author{Ruizhen Gu}
\orcid{0009-0001-8021-7052}
\affiliation{%
  \institution{Queen's University Belfast}
  \city{Belfast}
  \country{United Kingdom}
}
\email{r.gu@qub.ac.uk}

\author{Zhiwei Fei}
\orcid{0009-0009-3166-8525}
\affiliation{%
  \institution{Nanjing University}
  \city{Nanjing}
  \country{China}
}
\email{zhiweifei@smail.nju.edu.cn}

\author{Yazhuo Cao}
\orcid{0009-0002-1201-9908}
\affiliation{%
  \institution{King's College London}
  \city{London}
  \country{United Kingdom}
}
\email{yazhuo.cao@kcl.ac.uk}

\author{Lukas Twist}
\orcid{0009-0009-6640-2532}
\affiliation{%
  \institution{King's College London}
  \city{London}
  \country{United Kingdom}
}
\email{lukas.twist@kcl.ac.uk}

\author{Alina Geiger}
\orcid{0009-0002-3413-283X}
\affiliation{%
  \institution{Johannes Gutenberg University Mainz}
  \city{Mainz}
  \country{Germany}
}
\email{geiger@uni-mainz.de}

\author{Shuo Han}
\orcid{0009-0009-3413-3446}
\affiliation{%
  \institution{University College London}
  \city{London}
  \country{United Kingdom}
}
\email{shuo.han.25@ucl.ac.uk}

\author{Dominik Sobania}
\orcid{0000-0001-8873-7143}
\affiliation{%
  \institution{paluno, University of Duisburg-Essen}
  \city{Essen}
  \country{Germany}
}
\email{dominik.sobania@uni-due.de}

\author{Federica Sarro}
\orcid{0000-0002-9146-442X}
\affiliation{%
  \institution{University College London}
  \city{London}
  \country{United Kingdom}
}
\email{f.sarro@ucl.ac.uk}

\author{Jie M. Zhang}
\orcid{0000-0003-0481-7264}
\affiliation{%
  \institution{King's College London}
  \city{London}
  \country{United Kingdom}
}
\email{jie.zhang@kcl.ac.uk}

\renewcommand{\shortauthors}{Gong et al.}


\begin{abstract}
Agent skills are increasingly used to configure coding agents for software engineering (SE) tasks, yet current practice treats them as static, hand-crafted assets, or evolved on pass rate alone. This is insufficient: a skill can improve task success while substantially raising token cost, or introducing misleading guidance. We argue that SE agent skill bundles can be treated as multi-objective search objects and present \toolName{}, a framework that evolves skill bundles through LLM-proposed edits and NSGA-II Pareto selection on pass rate and inference cost.
Evaluated across all 16 SkillsBench SE tasks, \toolName{} achieves the top pass rate rank on 11 of 12 non-zero-pass tasks while achieving cost reductions of up to 32.1\% over static bundles, with pass rate gains up to 42 percentage points. Analysis of 38 skill edits shows that pruning and substitution dominate successful operations, offering actionable principles for skill bundle design.
Thereby, the current practice of deploying skills without cost-aware validation leaves better skill configurations unexplored, motivating a new class of multi-objective search-based skill engineering.
\end{abstract}


\maketitle

\setcounter{footnote}{0}

\section{Introduction}
\label{sec:intro}
Large language model (LLM)-based coding agents are increasingly configured through external artifacts: prompts, tools, memories, and \emph{agent skills} \cite{lee2026metaharness, gong2025tuning, li2026skillsbenchbenchmarkingagentskills}. Agent skills are portable folders that package instructions, scripts, and resources that an agent can load on demand for specialized work~\cite{anthropic2025introducingskills,AgentSkills}. In software engineering (SE), skills package reusable task-specific instructions, scripts, and resources for build, test, migration, and debugging workflows without retraining the model~\cite{trq2026claudecodeskills}.

Yet the emerging evidence is mixed. SkillsBench reports that skills improve performance overall and provide modest gains on SE tasks~\cite{li2026skillsbenchbenchmarkingagentskills}, while SWE-Skills-Bench finds that many public SWE skills provide no pass rate improvement and can introduce substantial token overhead~\cite{han2026swe}. Practitioners likewise report that skills require continuous refinement around concrete failure cases~\cite{trq2026claudecodeskills}. These findings suggest that a skill is not simply ``helpful'' or ``not helpful'': it creates a tradeoff between task success, inference cost, and contextual compatibility.

This motivates a search-based framing: \emph{agent skill bundles can be optimized and validated with these competing objectives}. Thereby, we propose \toolName{}, a multi-objective optimization (MOO) framework for SE agent skills. \toolName{} treats a skill bundle as a candidate solution, evaluates it with a task solver agent, and uses a skill optimizer agent to propose edits such as pruning, substitution, reordering, and rewriting. Candidate bundles are selected with NSGA-II using pass rate and cost as explicit objectives.

In summary, our main contributions are:
\begin{itemize}
    \item We argue that SE agent skills should be treated as multi-objective optimization objects, where pass rate and deployment cost are competing objectives rather than an afterthought, challenging the current literature of deploying and evaluating skills on pass rate alone.
    \item We propose \toolName{}, a MOO framework that combines LLM-proposed edits and NSGA-II survivor selection in a solver-optimizer loop.
    \item We provide empirical evidence on 16 SkillsBench SE tasks, demonstrating that static skill bundles leave Pareto-superior configurations unexplored.
    \item We analyze 38 filtered skill-edit patterns and identify pruning and substitution as the primary drivers of observed improvements, offering practical insights for skill bundle design.
    \item We release a replication package with data, scripts, and skill artifacts to support further research.
\end{itemize}


\section{Related Work}
SkillsBench~\cite{li2026skillsbenchbenchmarkingagentskills} introduced the first systematic paired evaluation of skills across diverse tasks, and SWE-Skills-Bench~\cite{han2026swe} found that many public SWE skills provide little benefit in SE tasks.
SkillRouter~\cite{zheng2026skillrouter} frames skill usage as a retrieval problem and shows that scalable deployment requires accurate routing over large skill pools.
Meanwhile, recent skill evolution approaches like EvoSkill~\cite{alzubi2026evoskillautomatedskilldiscovery}, Meta Context Engineering~\cite{ye2026metacontextengineeringagentic}, and EvoSkills~\cite{zhang2026evoskills} demonstrate that skills can themselves be optimized through iterative refinement, but they focus on pass rate alone.
Adjacent context and prompt optimization work, including ACE~\cite{zhang2025ace}, GEPA~\cite{agrawal2025gepa}, and Meta-Harness~\cite{lee2026metaharness}, optimizes contexts, prompts, or harness code around LLM systems. LLM-guided Genetic Improvement~\cite{evenmendoza2025llmguidedgi} is similar to our setting because it combines LLM semantic edits with search-based software evolution, but it does not optimize skill bundles.

Building on these, we target SE tasks with multi-objective search using LLM-proposed skill edits and NSGA-II survivor selection.

\begin{table*}[t]
    \centering
    \caption{Details of the 16 SE tasks used for evaluation.}
    \vspace{-0.3cm}
    {
    \setlength{\tabcolsep}{4pt}
    \resizebox{\textwidth}{!}{
    \begin{tabular}{r l l r r@{\hspace{0.55cm}}r l l r r}
    \toprule
    \textbf{ID} & \textbf{Task name} & \textbf{Category} & \textbf{\#Skills} & \textbf{\#Tests} &
    \textbf{ID} & \textbf{Task name} & \textbf{Category} & \textbf{\#Skills} & \textbf{\#Tests} \\
    \midrule
    \rowcolor{gray!20} 1 & \textsc{citation-check} & Info.\ Retrieval & 1 & 40 &
    9 & \textsc{jax-computing-basics} & Sci.\ Computing & 1 & 43 \\
    2 & \textsc{data-to-d3} & Data Visualization & 1 & 40 &
    10 & \textsc{parallel-tfidf-search} & Perf.\ Optimization & 3 & 40 \\
    \rowcolor{gray!20} 3 & \textsc{dialogue-parser} & NLP/Parsing & 1 & 40 &
    11 & \textsc{python-scala-translation} & Code Migration & 6 & 40 \\
    4 & \textsc{enterprise-information-search} & Info.\ Retrieval & 1 & 40 &
    12 & \textsc{react-performance-debugging} & Perf.\ Optimization & 2 & 40 \\
    \rowcolor{gray!20} 5 & \textsc{fix-build-agentops} & Build Repair & 8 & 40 &
    13 & \textsc{simpo-code-reproduction} & ML Reproduction & 2 & 40 \\
    6 & \textsc{fix-build-google-auto} & Build Repair & 3 & 40 &
    14 & \textsc{spring-boot-jakarta-migration} & Code Migration & 5 & 40 \\
    \rowcolor{gray!20} 7 & \textsc{flink-query} & Data Engineering & 2 & 41 &
    15 & \textsc{taxonomy-tree-merge} & Data Engineering & 1 & 40 \\
    8 & \textsc{gh-repo-analytics} & DevOps Analytics & 1 & 40 &
    16 & \textsc{trend-anomaly-causal-inference} & Data Analysis & 4 & 41 \\
    \bottomrule
    \end{tabular}
    }
    }
    \label{tab:task_overview}
    \end{table*}

\begin{figure}[t]
\centering
\includegraphics[width=\columnwidth]{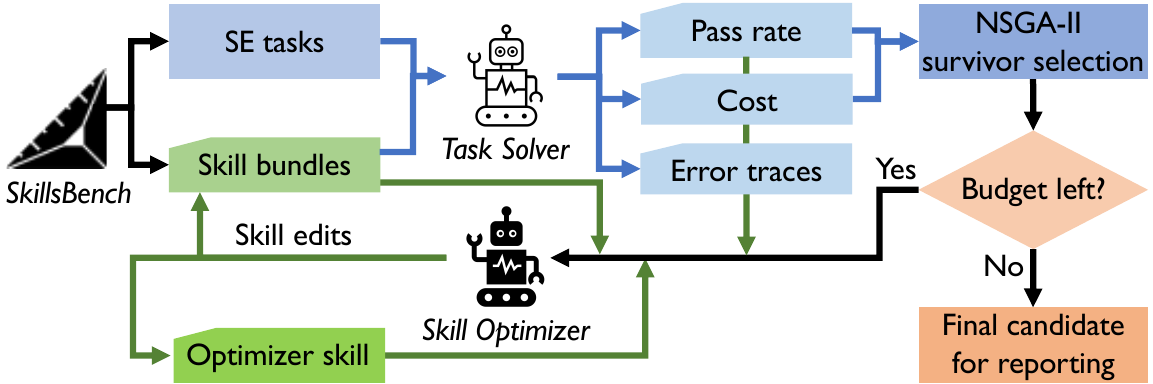}
\vspace{-0.5cm}
\caption{SkillMOO workflow: solver-optimizer loop with evolving skill bundles.}
\label{fig:workflow}
\end{figure}

\section{SkillMOO}
Specifically, \toolName{} uses a two-agent iterative workflow, as illustrated in \Cref{fig:workflow}. A \emph{candidate} is a task-specific skill bundle: a subset or edited version of the skills exposed to the coding agent. The search proceeds in three repeating phases:

\textbf{(1) Initialization.} Generation~0 seeds the population with candidates sampled as diverse subsets of the available skill pool. The \textbf{task solver agent} executes the SE task for each candidate and records pass rate, cost, runtime, and failure traces.

\textbf{(2) Optimization.} The \textbf{skill optimizer agent} receives each surviving bundle together with its evaluation evidence and decides what change is most likely to improve the pass/cost tradeoff, such as removing distracting guidance, replacing a misaligned skill, adding missing task-specific instructions, reordering exposed skills, or rewriting stale skill content. Each child is then evaluated by the same task solver.

\textbf{(3) Selection.} The current generation's evaluated candidates are ranked with NSGA-II~\cite{deb2002fast} non-dominated sorting and crowding-distance tie-breaking on the two objectives below; the top-ranked survivors breed the next generation's population.

\noindent Phases (2) and (3) repeat for each subsequent generation. The bi-objective formulation is:
\[
\min_{b \in \mathcal{B}} \; \mathbf{f}(b)=\left[-\text{pass}(b),\; \text{cost}(b)\right]
\]
where $\text{pass}(b)$ is verifier pass rate and $\text{cost}(b)$ is LLM inference cost. This distinguishes \toolName{} from skill discovery or routing: it asks not only which skill helps, but which bundle offers the best pass/cost tradeoff.

\section{Experimental Setup}

We evaluate all 16 SE tasks in SkillsBench~\cite{li2026skillsbenchbenchmarkingagentskills}, summarized in \Cref{tab:task_overview}. Each method is run 10 times per task; a run passes only when its solution passes the full automated test suite, so pass rate is the fraction of passing runs. Because default verifiers contain roughly 3--10 tests, we expanded every suite to 40--43 tests with GPT-5.4: behavior checks cover required functionality and safety checks reject invalid or unsafe outputs while retaining original compile/build gates. Each added test was checked manually against the task specification and reference environment, generated without skill content, and applied uniformly across methods.

We use \texttt{GLM-5}~\cite{zeng2026glm} for both task solver and skill optimizer agents. The evaluation uses population size 4 and 3 generations, which are chosen to bound optimization budget: with per-candidate evaluation costs of \$0.1--\$14 (USD, at GLM-5 API pricing at time of evaluation) and a 900 s timeout, this configuration keeps per-task optimization overhead practical (\$1--\$15 USD and under three hours). Scott-Knott Effect Size Difference (ESD) pass rate ranks~\cite{scott1974cluster} are computed from the 10-run vectors across all tasks.

RQ2 hypervolume (HV) is computed after min-max normalizing pass rate and cost per-task across all evaluated candidates; the reference point is $(0,\,1.1)$ in normalized space (worst pass, 10\% beyond worst cost). RQ3 uses only edits drawn from the Pareto archive of fully evaluated candidates, applying two inclusion criteria: (1) the bundle change is meaningful; and (2) pass rate, cost, and runtime deltas against the \texttt{ori\_skill} baseline are non-zero.

We evaluate \toolName{}, \texttt{ori\_skill} (original skill bundle), and \texttt{no\_skill} (no guidance)
, addressing three research questions (RQs):
\begin{description}
    \item[\ding{228} RQ1.] How does \toolName{} compare with baselines on pass rate and cost?
    \item[\ding{228} RQ2.] Is \toolName{} optimization economically worthwhile accounting for overhead?
    \item[\ding{228} RQ3.] What do skill-edit patterns reveal about optimized skills for SE tasks?
\end{description}
RQ1 establishes effectiveness, RQ2 determines whether the search cost is recoverable through skill reuse, and RQ3 examines which edits are associated with improvements.

\section{Results and Analysis}
\label{sec:results}

\subsection{RQ1: Effectiveness of \toolName{}}
\Cref{tab:rq1_final_results} reports all 16 SkillsBench SE tasks with \texttt{GLM-5}. In particular, \toolName{} achieves the top pass rate rank on 11 of 12 non-zero-pass tasks (4 tasks yield zero pass for all methods\footnote{Inspection of execution logs shows that these four tasks share heavy or specialized runtime dependencies that stress container memory and build-time budgets, causing the agent to fail, therefore we exclude them from comparative analysis.}); the strongest gain on a non-trivial task is on \textsc{Task 5}: pass rate rises from 0.16 to 0.37 (+21.3 percentage points) while cost falls by 32.1\% and runtime by 23.6\%, with Scott-Knott ESD confirming statistical significance.

Notably, \texttt{no\_skill} achieves rank 1 cost on 12 of 16 tasks and has the lowest mean cost on every task — as expected, since it incurs no skill-content context overhead — though on four tasks its cost is not statistically distinct from another method's under Scott-Knott ESD, given run-to-run cost variance. \texttt{no\_skill} still trails SkillMOO on pass rate for 11 of 12 non-zero-pass tasks. However, \texttt{no\_skill} leads pass rate only on \textsc{gh-repo-analytics}, and several tasks yield near-zero pass for all methods. Thus, selectively pruning a peripheral skill can help while removing all skills discards relevant task guidance, motivating multi-objective optimization of the pass/cost tradeoff.

\vspace{0.1cm}
\begin{quotebox}
\textbf{RQ1 answer:} Across all 16 tasks, \toolName{} achieves the top pass rate rank on 11 of 12 non-zero-pass tasks. On tasks with larger skill bundles, \toolName{} consistently finds better pass/cost tradeoffs than static deployment, with pass rate gains up to +42 percentage points and cost reductions up to 32.1\%.
\end{quotebox}


\begin{table*}[t]
\centering
\caption{Effectiveness and cost comparison on all 16 SkillsBench SE tasks. Pass rate and cost are mean $\pm$ SD over 10 independent runs. \#Skills is the mean number of skills in the selected bundle across runs (fractional values arise from run-to-run variation in the selected candidate). $r_{p}$/$r_{c}$: Scott-Knott ESD rank on pass rate / cost rank (lower cost = better $r_{c}$).}
\vspace{-0.3cm}
{
\tiny
\renewcommand{\arraystretch}{0.85}
\setlength{\tabcolsep}{4pt}
\adjustbox{width=\textwidth}{
\begin{tabular}{c c c c c c c c c c c c c c c c}
\toprule
\multirow{2}{*}{\textbf{Task ID}} &
\multicolumn{5}{c}{\textbf{\toolName{}}} &
\multicolumn{5}{c}{\textbf{\texttt{ori\_skill}}} &
\multicolumn{5}{c}{\textbf{\texttt{no\_skill}}} \\
\cmidrule(lr){2-6}\cmidrule(lr){7-11}\cmidrule(lr){12-16}
& \textbf{\#Skills} & \textbf{$r_{p}$} & \textbf{Pass Rate} & \textbf{$r_{c}$} & \textbf{Cost} &
\textbf{\#Skills} & \textbf{$r_{p}$} & \textbf{Pass Rate} & \textbf{$r_{c}$} & \textbf{Cost} &
\textbf{\#Skills} & \textbf{$r_{p}$} & \textbf{Pass Rate} & \textbf{$r_{c}$} & \textbf{Cost} \\
\midrule
1 & 1.0 & \cellcolor{frenchblue!35}\textbf{1} & \cellcolor{frenchblue!35}\textbf{0.90$\pm$0.32} & \cellcolor{dollarbill!35}\textbf{1} & \cellcolor{dollarbill!35}\textbf{0.25$\pm$0.24} &
1.0 & \cellcolor{frenchblue!12}3 & \cellcolor{frenchblue!12}0.10$\pm$0.32 & \cellcolor{dollarbill!24}2 & \cellcolor{dollarbill!24}0.31$\pm$0.07 &
0.0 & \cellcolor{frenchblue!24}2 & \cellcolor{frenchblue!24}0.60$\pm$0.52 & \cellcolor{dollarbill!24}2 & \cellcolor{dollarbill!24}0.22$\pm$0.25 \\
2 & 1.0 & \cellcolor{frenchblue!35}\textbf{1} & \cellcolor{frenchblue!35}\textbf{0.89$\pm$0.03} & \cellcolor{dollarbill!35}\textbf{1} & \cellcolor{dollarbill!35}\textbf{0.25$\pm$0.11} &
1.0 & \cellcolor{frenchblue!12}3 & \cellcolor{frenchblue!12}0.16$\pm$0.34 & \cellcolor{dollarbill!24}2 & \cellcolor{dollarbill!24}0.31$\pm$0.14 &
0.0 & \cellcolor{frenchblue!24}2 & \cellcolor{frenchblue!24}0.33$\pm$0.42 & \cellcolor{dollarbill!24}2 & \cellcolor{dollarbill!24}0.22$\pm$0.21 \\
3 & 1.0 & \cellcolor{frenchblue!35}\textbf{1} & \cellcolor{frenchblue!35}\textbf{0.79$\pm$0.28} & \cellcolor{dollarbill!24}2 & \cellcolor{dollarbill!24}0.27$\pm$0.18 &
1.0 & \cellcolor{frenchblue!35}\textbf{1} & \cellcolor{frenchblue!35}\textbf{0.35$\pm$0.46} & \cellcolor{dollarbill!10}3 & \cellcolor{dollarbill!10}0.33$\pm$0.19 &
0.0 & \cellcolor{frenchblue!24}2 & \cellcolor{frenchblue!24}0.35$\pm$0.45 & \cellcolor{dollarbill!35}\textbf{1} & \cellcolor{dollarbill!35}\textbf{0.24$\pm$0.08} \\
4 & 1.0 & \cellcolor{frenchblue!35}\textbf{1} & \cellcolor{frenchblue!35}\textbf{0.45$\pm$0.37} & \cellcolor{dollarbill!35}\textbf{1} & \cellcolor{dollarbill!35}\textbf{0.33$\pm$0.23} &
1.0 & \cellcolor{frenchblue!12}3 & \cellcolor{frenchblue!12}0.00$\pm$0.00 & \cellcolor{dollarbill!35}\textbf{1} & \cellcolor{dollarbill!35}\textbf{0.41$\pm$0.00} &
0.0 & \cellcolor{frenchblue!24}2 & \cellcolor{frenchblue!24}0.23$\pm$0.36 & \cellcolor{dollarbill!35}\textbf{1} & \cellcolor{dollarbill!35}\textbf{0.32$\pm$0.77} \\
5 & 4.0 & \cellcolor{frenchblue!35}\textbf{1} & \cellcolor{frenchblue!35}\textbf{0.37$\pm$0.16} & \cellcolor{dollarbill!35}\textbf{1} & \cellcolor{dollarbill!35}\textbf{1.10$\pm$0.33} &
8.0 & \cellcolor{frenchblue!24}2 & \cellcolor{frenchblue!24}0.16$\pm$0.08 & \cellcolor{dollarbill!24}2 & \cellcolor{dollarbill!24}1.61$\pm$0.41 &
0.0 & \cellcolor{frenchblue!12}3 & \cellcolor{frenchblue!12}0.10$\pm$0.00 & \cellcolor{dollarbill!24}2 & \cellcolor{dollarbill!24}1.06$\pm$0.71 \\
6 & 1.6 & \cellcolor{frenchblue!35}\textbf{1} & \cellcolor{frenchblue!35}\textbf{0.42$\pm$0.25} & \cellcolor{dollarbill!24}2 & \cellcolor{dollarbill!24}0.17$\pm$0.42 &
3.0 & \cellcolor{frenchblue!24}2 & \cellcolor{frenchblue!24}0.00$\pm$0.00 & \cellcolor{dollarbill!24}2 & \cellcolor{dollarbill!24}0.19$\pm$0.02 &
0.0 & \cellcolor{frenchblue!24}2 & \cellcolor{frenchblue!24}0.00$\pm$0.00 & \cellcolor{dollarbill!35}\textbf{1} & \cellcolor{dollarbill!35}\textbf{0.13$\pm$0.01} \\
7 & 1.3 & \cellcolor{frenchblue!35}\textbf{1} & \cellcolor{frenchblue!35}\textbf{0.00$\pm$0.00} & \cellcolor{dollarbill!24}2 & \cellcolor{dollarbill!24}1.14$\pm$0.00 &
2.0 & \cellcolor{frenchblue!35}\textbf{1} & \cellcolor{frenchblue!35}\textbf{0.00$\pm$0.00} & \cellcolor{dollarbill!10}3 & \cellcolor{dollarbill!10}1.42$\pm$0.02 &
0.0 & \cellcolor{frenchblue!35}\textbf{1} & \cellcolor{frenchblue!35}\textbf{0.00$\pm$0.00} & \cellcolor{dollarbill!35}\textbf{1} & \cellcolor{dollarbill!35}\textbf{1.01$\pm$0.01} \\
8 & 1.0 & \cellcolor{frenchblue!24}2 & \cellcolor{frenchblue!24}0.16$\pm$0.30 & \cellcolor{dollarbill!35}\textbf{1} & \cellcolor{dollarbill!35}\textbf{0.19$\pm$0.18} &
1.0 & \cellcolor{frenchblue!24}2 & \cellcolor{frenchblue!24}0.15$\pm$0.30 & \cellcolor{dollarbill!24}2 & \cellcolor{dollarbill!24}0.24$\pm$0.24 &
0.0 & \cellcolor{frenchblue!35}\textbf{1} & \cellcolor{frenchblue!35}\textbf{0.22$\pm$0.35} & \cellcolor{dollarbill!24}2 & \cellcolor{dollarbill!24}0.17$\pm$0.16 \\
9 & 1.0 & \cellcolor{frenchblue!35}\textbf{1} & \cellcolor{frenchblue!35}\textbf{0.28$\pm$0.45} & \cellcolor{dollarbill!24}2 & \cellcolor{dollarbill!24}0.25$\pm$0.11 &
1.0 & \cellcolor{frenchblue!12}3 & \cellcolor{frenchblue!12}0.00$\pm$0.00 & \cellcolor{dollarbill!10}3 & \cellcolor{dollarbill!10}0.32$\pm$0.00 &
0.0 & \cellcolor{frenchblue!24}2 & \cellcolor{frenchblue!24}0.19$\pm$0.40 & \cellcolor{dollarbill!35}\textbf{1} & \cellcolor{dollarbill!35}\textbf{0.23$\pm$0.09} \\
10 & 1.7 & \cellcolor{frenchblue!35}\textbf{1} & \cellcolor{frenchblue!35}\textbf{0.10$\pm$0.32} & \cellcolor{dollarbill!24}2 & \cellcolor{dollarbill!24}0.84$\pm$0.19 &
3.0 & \cellcolor{frenchblue!24}2 & \cellcolor{frenchblue!24}0.00$\pm$0.00 & \cellcolor{dollarbill!10}3 & \cellcolor{dollarbill!10}1.06$\pm$0.00 &
0.0 & \cellcolor{frenchblue!24}2 & \cellcolor{frenchblue!24}0.10$\pm$0.31 & \cellcolor{dollarbill!35}\textbf{1} & \cellcolor{dollarbill!35}\textbf{0.75$\pm$0.34} \\
11 & 2.0 & \cellcolor{frenchblue!35}\textbf{1} & \cellcolor{frenchblue!35}\textbf{0.51$\pm$0.12} & \cellcolor{dollarbill!24}2 & \cellcolor{dollarbill!24}1.06$\pm$0.25 &
6.0 & \cellcolor{frenchblue!35}\textbf{1} & \cellcolor{frenchblue!35}\textbf{0.39$\pm$0.06} & \cellcolor{dollarbill!24}2 & \cellcolor{dollarbill!24}1.12$\pm$0.36 &
0.0 & \cellcolor{frenchblue!24}2 & \cellcolor{frenchblue!24}0.39$\pm$0.16 & \cellcolor{dollarbill!35}\textbf{1} & \cellcolor{dollarbill!35}\textbf{0.85$\pm$0.31} \\
12 & 1.4 & \cellcolor{frenchblue!35}\textbf{1} & \cellcolor{frenchblue!35}\textbf{0.36$\pm$0.29} & \cellcolor{dollarbill!24}2 & \cellcolor{dollarbill!24}0.88$\pm$0.50 &
2.0 & \cellcolor{frenchblue!24}2 & \cellcolor{frenchblue!24}0.01$\pm$0.02 & \cellcolor{dollarbill!10}3 & \cellcolor{dollarbill!10}1.11$\pm$0.14 &
0.0 & \cellcolor{frenchblue!12}3 & \cellcolor{frenchblue!12}0.00$\pm$0.00 & \cellcolor{dollarbill!35}\textbf{1} & \cellcolor{dollarbill!35}\textbf{0.79$\pm$0.01} \\
13 & 1.4 & \cellcolor{frenchblue!35}\textbf{1} & \cellcolor{frenchblue!35}\textbf{0.00$\pm$0.00} & \cellcolor{dollarbill!24}2 & \cellcolor{dollarbill!24}1.14$\pm$0.01 &
2.0 & \cellcolor{frenchblue!35}\textbf{1} & \cellcolor{frenchblue!35}\textbf{0.00$\pm$0.00} & \cellcolor{dollarbill!10}3 & \cellcolor{dollarbill!10}1.42$\pm$0.01 &
0.0 & \cellcolor{frenchblue!35}\textbf{1} & \cellcolor{frenchblue!35}\textbf{0.00$\pm$0.00} & \cellcolor{dollarbill!35}\textbf{1} & \cellcolor{dollarbill!35}\textbf{1.01$\pm$0.02} \\
14 & 5.0 & \cellcolor{frenchblue!35}\textbf{1} & \cellcolor{frenchblue!35}\textbf{0.99$\pm$0.01} & \cellcolor{dollarbill!24}2 & \cellcolor{dollarbill!24}1.25$\pm$0.49 &
5.0 & \cellcolor{frenchblue!24}2 & \cellcolor{frenchblue!24}0.97$\pm$0.00 & \cellcolor{dollarbill!10}3 & \cellcolor{dollarbill!10}1.55$\pm$0.27 &
0.0 & \cellcolor{frenchblue!12}3 & \cellcolor{frenchblue!12}0.91$\pm$0.01 & \cellcolor{dollarbill!35}\textbf{1} & \cellcolor{dollarbill!35}\textbf{1.14$\pm$0.23} \\
15 & 1.0 & \cellcolor{frenchblue!35}\textbf{1} & \cellcolor{frenchblue!35}\textbf{0.00$\pm$0.00} & \cellcolor{dollarbill!24}2 & \cellcolor{dollarbill!24}1.14$\pm$0.04 &
1.0 & \cellcolor{frenchblue!35}\textbf{1} & \cellcolor{frenchblue!35}\textbf{0.00$\pm$0.00} & \cellcolor{dollarbill!10}3 & \cellcolor{dollarbill!10}1.42$\pm$0.03 &
0.0 & \cellcolor{frenchblue!35}\textbf{1} & \cellcolor{frenchblue!35}\textbf{0.00$\pm$0.00} & \cellcolor{dollarbill!35}\textbf{1} & \cellcolor{dollarbill!35}\textbf{1.01$\pm$0.02} \\
16 & 1.9 & \cellcolor{frenchblue!35}\textbf{1} & \cellcolor{frenchblue!35}\textbf{0.00$\pm$0.00} & \cellcolor{dollarbill!24}2 & \cellcolor{dollarbill!24}1.14$\pm$0.02 &
4.0 & \cellcolor{frenchblue!35}\textbf{1} & \cellcolor{frenchblue!35}\textbf{0.00$\pm$0.00} & \cellcolor{dollarbill!10}3 & \cellcolor{dollarbill!10}1.42$\pm$0.03 &
0.0 & \cellcolor{frenchblue!35}\textbf{1} & \cellcolor{frenchblue!35}\textbf{0.00$\pm$0.00} & \cellcolor{dollarbill!35}\textbf{1} & \cellcolor{dollarbill!35}\textbf{1.01$\pm$0.03} \\
\midrule
\textbf{Avg} & 1.7 & \cellcolor{frenchblue!35}\textbf{1.1} & \cellcolor{frenchblue!35}\textbf{0.39} & \cellcolor{dollarbill!24}{1.7} & \cellcolor{dollarbill!24}{0.71} &
2.6 & \cellcolor{frenchblue!24}{1.9} & \cellcolor{frenchblue!24}{0.14} & \cellcolor{dollarbill!10}{2.5} & \cellcolor{dollarbill!10}{0.89} &
0.0 & \cellcolor{frenchblue!24}{1.9} & \cellcolor{frenchblue!24}{0.21} & \cellcolor{dollarbill!35}\textbf{1.3} & \cellcolor{dollarbill!35}\textbf{0.64} \\
\bottomrule
\end{tabular}
}
}
\label{tab:rq1_final_results}
\end{table*}

\begin{table*}[t]
\centering
\caption{\toolName{} optimization overhead (USD), two-objective HV improvement ($\Delta$HV (\%)), and optimization efficiency (Cost/$\Delta$HV\%) for the 12 tasks with non-zero pass rate.}
\vspace{-0.3cm}
\label{tab:rq2_one_shot_value}
{
\renewcommand{\arraystretch}{0.9}
\setlength{\tabcolsep}{3pt}
\resizebox{\textwidth}{!}{
\begin{tabular}{r r r r r r r@{\hspace{0.45cm}}r r r r r r r}
\toprule
\textbf{ID} & \textbf{Opt. Cost} & \textbf{Opt. Runtime} & \textbf{HV} & \textbf{HV} & \textbf{$\Delta$HV} & \textbf{Cost/$\Delta$HV\%} &
\textbf{ID} & \textbf{Opt. Cost} & \textbf{Opt. Runtime} & \textbf{HV} & \textbf{HV} & \textbf{$\Delta$HV} & \textbf{Cost/$\Delta$HV\%} \\
& \textbf{(USD)} & \textbf{(s)} & \textbf{(\texttt{ori\_skill})} & \textbf{(\toolName{})} & \textbf{(\%)} & &
& \textbf{(USD)} & \textbf{(s)} & \textbf{(\texttt{ori\_skill})} & \textbf{(\toolName{})} & \textbf{(\%)} & \\
\midrule
\rowcolor{gray!20} 1 & 4.1539 & 9720 & 0.0016 & 0.0798 & 5031 & 0.0008 &
 8 & 1.9664 &  832 & 0.0012 & 0.0044 &  263 & 0.0075 \\
2 & 4.1437 & 4613 & 0.0015 & 0.0723 & 4580 & 0.0009 &
 9 & 2.8053 & 1717 & 0.0016 & 0.0319 & 1903 & 0.0015 \\
\rowcolor{gray!20} 3 & 3.3556 & 6428 & 0.0017 & 0.0489 & 2828 & 0.0012 &
10 & 9.6070 & 5083 & 0.0053 & 0.0480 &  807 & 0.0119 \\
4 & 4.2755 & 5191 & 0.0020 & 0.0612 & 2892 & 0.0015 &
11 & 1.8611 & 1545 & 0.0056 & 0.0296 &  430 & 0.0043 \\
\rowcolor{gray!20} 5 & 2.2676 & 2305 & 0.0081 & 0.1783 & 2110 & 0.0011 &
12 & 9.9575 & 6233 & 0.0055 & 0.1342 & 2322 & 0.0043 \\
6 & 13.7338 & 3998 & 0.0009 & 0.0183 & 1845 & 0.0074 &
14 & 1.7641 &  867 & 0.0078 & 0.0311 &  301 & 0.0059 \\
\bottomrule
\end{tabular}
}
}
\end{table*}

\subsection{RQ2: Optimization Efficiency}
\Cref{tab:rq2_one_shot_value} reports optimization cost and resulting hypervolume (HV) uplift over \texttt{ori\_skill} for the 12 tasks with non-zero pass rate. HV is computed in a per-task normalized two-objective space. Because percentage HV gains can be large when \texttt{ori\_skill} starts near zero, we interpret RQ2 primarily through optimization cost and reuse efficiency rather than percentage uplift alone. 

Overall, \toolName{} improves HV on all 12 reported tasks, but the overhead varies substantially: optimization costs range from \$1.76 to \$13.73. Dividing this one-time cost by the observed per-run saving over \texttt{ori\_skill} yields an approximate cost-only break-even of 5--682 reuses (median 45); a single-shot task instead requires a reliability gain worth the search cost. Although all search runs are paid before selection, the Pareto archive retains alternatives for a deployment budget or reliability target rather than simply choosing the highest-pass run.

\vspace{0.1cm}
\begin{quotebox}
\textbf{RQ2 answer:} 
On the 12 non-zero-pass tasks, \toolName{} search costs between \$1.76 and \$13.73 each, and each one-percent HV gain costs well under two U.S.\ cents in search fees---so the overhead stays worthwhile relative to typical inference spendings.
\end{quotebox}

\begin{table*}[t!]
\centering
\caption{Exploratory pattern analysis of filtered skill-bundle operations against the corresponding \texttt{ori\_skill} baseline.}
\vspace{-0.3cm}
{
\renewcommand{\arraystretch}{0.85}
\setlength{\tabcolsep}{10pt}
\resizebox{\textwidth}{!}{
\begin{tabular}{llcccc}
\toprule
\textbf{Operation} & \textbf{Description} & \textbf{\#Edits} & \textbf{Pass rate$\uparrow$} & \textbf{Cost$\downarrow$} & \textbf{Runtime$\downarrow$} \\
\midrule
\multirow{2}{*}{Remove Skills}
  & Drop peripheral or misaligned skills from bundle; observed with pass/cost gain & 11 & 8/11 & 9/11 & 5/11 \\
  & Remove one skill indicated as redundant by evaluation or test evidence        & 5  & 5/5  & 2/5  & 2/5  \\
\midrule
Replace Skills
  & Replace lower-priority skills with alternatives selected by the optimizer & 7 & 5/7 & 7/7 & 4/7 \\
\midrule
\multirow{3}{*}{Add Skills}
  & Add skills inferred as missing from current bundle       & 5 & 0/5 & 5/5 & 5/5 \\
  & Explicitly add object-oriented programming guidance      & 3 & 0/3 & 3/3 & 3/3 \\
  & Explicitly add functional programming style guidance     & 1 & 0/1 & 1/1 & 1/1 \\
\midrule
\multirow{2}{*}{Edit Skill Content}
  & Rewrite skill to remove lingering deprecated API mentions & 2 & 2/2 & 2/2 & 1/2 \\
  & Remove never-triggered guidance sections within a skill  & 2 & 0/2 & 2/2 & 2/2 \\
\midrule
\multirow{2}{*}{Reorder Bundle}
  & Promote foundational setup step first to prevent timeout  & 1 & 1/1 & 1/1 & 0/1 \\
  & Reorder bundle to reflect optimal invocation sequence     & 1 & 0/1 & 1/1 & 1/1 \\
\bottomrule
\end{tabular}
}
}
\label{tab:rq3_pattern_evidence}
\end{table*}

\subsection{RQ3: Skill-Edit Pattern Evidence}
\Cref{tab:rq3_pattern_evidence} shows that Remove Skills and Replace Skills are more often associated with pass rate improvements than Add Skills in the filtered archive. Remove Skills is the most frequent operation with 16 edits: dropping peripheral or misaligned skills (11 edits, 8/11 pass↑, 9/11 cost↓) and removing a single redundant skill (5 edits, 5/5 pass↑). Add Skills shows 0/9 pass improvements across its three sub-types, which is only suggestive evidence that adding guidance did not help in these logged cases. Edit Skill Content is the most targeted operation: removing outdated references yields 2/2 pass improvements, while pruning unused sections reduces cost without affecting pass rate. These patterns suggest that the observed SkillMOO gains often come from removing irrelevant guidance and replacing misaligned content, rather than adding new instructions---mirroring the RQ1 finding that pass improvements arise only when edits address task-specific verifier failures. \footnote{Because the edits are filtered from search traces, RQ3 should be read as exploratory pattern evidence rather than an independent causal study.}

\vspace{0.1cm}
\begin{quotebox}
\textbf{RQ3 answer:} In the filtered edit archive, pruning and substitution operations are most often associated with successful edits (16 and 7 edits respectively); these descriptive patterns motivate, but do not prove, general rules for skill design.
\end{quotebox}

\subsection{Threats to Validity}
First, our evaluation covers a single benchmark, with 1--8 initial skills; larger skill pools and repositories may incur higher search and evaluation costs.
Second, test suites are expanded to 40--43 tests via LLM generation; while the expanded tests were generated without access to skill content and validated manually, they were produced by the same model family used for task solving, which may introduce shared reasoning biases in borderline cases — compile and build gates are preserved as model-independent ground truth to mitigate this.
Third, we report results for one LLM (\texttt{GLM-5}), selected for its competitive cost-to-capability ratio and API availability at evaluation time; whether the MOO framing and Pareto improvements transfer to other coding agents (e.g., GPT or Claude) is an open empirical question that future work should address.

\section{Conclusion and Research Agenda}

Overall, our findings suggest that effective SE agent skills should be evaluated and evolved for both task success and inference cost, rather than treated as static assets or optimized for pass rate alone, motivating skill engineering as a new SBSE research direction.

\para{Research Agenda}
Specifically, we suggest four future directions:

\begin{itemize}
    \item \textbf{Causal evaluation of skill edits:} Our edit-pattern evidence is descriptive and drawn from filtered search traces. Future work should isolate edit operations through controlled ablations, randomized edit proposals, and repeated validation to determine when an edit causally improves agent behavior.
    \item \textbf{Cost-aware skill routing and reuse:} Running optimization per task can be expensive. A natural next step is to reuse learned Pareto fronts or edit histories to route future tasks to existing skill configurations under a user-specified budget, avoiding full re-optimization at inference time.
    \item \textbf{Cross-task and cross-model transfer:} The same skill bundle may interact differently with task type, verifier structure, and underlying LLM. Future work should test whether optimized bundles, edit heuristics, or routing policies transfer across unseen SE tasks and across different coding agents.
    \item \textbf{Continuous skill maintenance:} Skills can become stale as APIs, dependencies, models, and project conventions evolve. Lightweight SkillMOO-style revalidation could support continuous maintenance, detecting when a previously useful skill becomes costly, misleading, or obsolete.
\end{itemize}

\para{Acknowledgment}
This work has been supported by the ITEA GENIUS grant (project number 23026).

\section*{Data Availability Statement}
The replication package, including existing evaluation records, per-task and per-method aggregation outputs, source code, and scripts to rebuild all reported tables, is available at \textcolor{blue}{\url{https://github.com/gjz78910/SkillMOO}}.

\bibliographystyle{ACM-Reference-Format}
\bibliography{references}

\end{document}